\documentstyle[rotate,aps,preprint,psfig]{revtex}

\def\beq{\begin{equation}}
\def\eeq{\end{equation}}
\def\bea{\begin{eqnarray}}
\def\eea{\end{eqnarray}}
\def\NL{\nonumber \\}
\def\dag{\dagger}

\begin{document}

\draft

\title{Damping rate and Lyapunov exponent of a Higgs field at high temperature}
\author{T.S. Bir\'o$^1$ and M.H. Thoma$^2$}
\address{$^1$MTA KFKI RMKI Theory Division,\\ H-1525 Budapest P.O.Box 49, Hungary}
\address{$^2$Institut f\"ur Theoretische Physik, Universit\"at Giessen,\\
D-35392 Giessen, Germany}
\date{\today}
\maketitle

\begin {abstract}

The damping rate of a Higgs field at zero momentum is calculated using
the Braaten-Pisarski method and compared to the Lyapunov exponent
of the classical SU(2) Yang-Mills Higgs system.

\end{abstract}

\pacs{PACS numbers: 11.15.Ha,12.90.+b}


\section{Introduction}

Since the advent of gauge theories in the 1970-s there is a continous 
need for their application at high temperatures characteristic to
high energy collisions of heavy nuclei or the early universe.
Even asymptotically free theories, like QCD, prove to be far from
trivial at high energy density or temperature.

Rather recently a promising method has been invented, the hard thermal loop
(HTL) resummation \cite{BP1}, for treating field theories at high temperature
consistently.
This method is analytical, complementing extended numerical simulations
on a space time lattice at finite temperature.

One of the most famous results of the HTL method, the
gluon damping rate at zero momentum \cite{BP2}, was found to coincide with
half of the Lyapunov exponent of a classical real-time lattice
description of nonabelian gluon fields \cite{BGMT}. An attempt for
understanding
this at the first sight rather surprising agreement between two different
quantities of high temperature gluon dynamics has been recently made
by emphasizing that the main contribution to the gluon damping rate
of order $g^2T$ stems from long wavelength gluons (i.e. from the
infrared part of the gluon distribution) and hence can be a measure
of the self-ergodizing speed of classically chaotic gluon fields \cite{BGM95}.

In the present article we investigate whether a similar correspondence
can be found in the coupled Yang-Mills Higgs-doublet system between
the Higgs damping rate calculated in the HTL approximation and the
Lyapunov exponent calculated on the lattice.
If yes, it would let us surmise that something inherently classical
is contained in the HTL method in the order of $g^2T$ and their
results in principle could be simulated by classical dynamical
calculations. If not, it would warn us that there is something uniquely
particular for the pure gauge field systems.
In fact chaos is opposite to damping, but because of
time symmetry exponential growth and damping can have the same rate
in given approximations. 
These are two facets of the same dynamics studied by using different
boundary conditions in time.

Although the Yang-Mills-Higgs system has already
been studied classically and found to be chaotic \cite{BGMT}, for a throughout
comparison with the new HTL results presented here we have to redo
and extend that calculation, too.

\section{Damping rates}

The aim of this section is the calculation of the Higgs and gauge boson
damping rates at zero momentum and high temperature for comparing it with
the Lyapunov exponent of the classical SU(2) Yang-Mills system coupled
to a Higgs field \cite{BGMT}. We start from the Lagrangian
\beq
{\cal L}=-\frac{1}{4}\> F_{\mu \nu}^a\, F^{\mu \nu}_a+(D_\mu \Phi )^\dagger \,
(D^\mu \Phi )+\mu ^2 \> \Phi ^\dagger \, \Phi -\lambda \> (\Phi ^\dagger \,
\Phi )^2,
\label{e1}
\eeq
where $\Phi $ is a charged Higgs doublet and $D_\mu \equiv \partial _\mu
-g\, A_a^\mu T^a$ the covariant derivative containing the SU(2) generators
$T^a$. At high temperatures above the phase transition 
$\Phi^{\dag}\Phi \sim T^2$ and the quadratic term can be neglected besides
the quartic term. Throughout this paper we consider $\mu = 0$.

The damping rates follow from the imaginary part of the Higgs and gauge
boson self energies, respectively. In order to obtain gauge invariant
results, complete to leading order in the coupling constants, we adopt
the Braaten-Pisarski method \cite{BP1}. It is based on the distinction
between hard momenta of the order of the temperature $T$ and soft ones
of the order $gT$ or ${\sqrt \lambda }T$. The Braaten-Pisarski method
amounts to the use of effective Green functions constructed by resumming
the HTL diagrams. Calculating the gluon
damping rate at zero momentum in this way solved the plasmon puzzle,
a famous problem of finite temperature gauge theories \cite{BP2}.

The first step is to extract the HTL contributions to the effective
Green functions, which arise from one-loop diagrams where the internal
momenta are hard. Assuming all external momenta of the Green functions to be
soft, the HTL corrections are of the same order in $g$ and $\lambda $
as the bare Green functions. In the case of a Yang-Mills field coupled
to the Higgs field described by the Lagrangian (\ref{e1}) there are no
HTL corrections to effective vertices involving Higgs lines, as can
be shown by power counting \cite{BP1}. This observation holds also for scalar
QED \cite{KRS}, which is apart from the gauge boson self coupling and
SU(2) factors identical to (\ref{e1}), as well as
for the Yukawa theory \cite{THO}.

The HTL Higgs self energy, from which the effective Higgs propagator
is constructed by resummation, follows from the diagrams of Fig.1, where
the internal momenta of the polarization graph are hard. Adopting the
usual Feynman rules \cite{IZ}, taking into account symmetry factors 1/2
for the tadpole diagrams, we obtain the HTL Higgs self energy using
the Matsubara formalism,
\beq
\Xi _{\alpha \beta} =\delta _{\alpha \beta}\> \left(\frac{4}{3}\, \lambda
+\frac{3}{16} g^2\right)  T^2
\label{e2}
\eeq
with the SU(2) indices $\alpha $,$\beta $ of the fundamental representation.
As in the case of scalar QED \cite{KRS} and the Yukawa theory \cite{THO}
the HTL self energy is momentum independent and real.

The effective Higgs propagator is given by resumming the HTL self energy
(\ref{e2}) in a Dyson-Schwinger equation leading to
\beq
\Delta ^\star _{\alpha \beta}=\frac{\delta _{\alpha \beta}}{K^2-m_H^2},
\label{e3}
\eeq
where $K^2=k_0^2-k^2$, $k\equiv |{\bf k}|$ and $m_H^2=(4\lambda /3+3g^2/16)\,
T^2$ is the square of the thermal Higgs mass.

The HTL gauge boson self energy is shown in Fig.2. The first three diagrams,
containing the gauge boson self couplings and a ghost loop, give the
well known results for the longitudinal and transverse parts
of the self energies \cite{WEL1,KLI}
\bea
\Pi _L^{ab}(K) & = & -3\, \delta ^{ab} \, m_g^2\> \left (1-\frac{k_0}{2k}\,
\ln \frac {k_0+k}{k_0-k} \right ), \nonumber \\
\Pi _T^{ab}(K) & = & \frac{3}{2}\, \delta ^{ab}\, m_g^2\, \frac{k_0^2}{k^2}\>
\left [1-\left (1-\frac{k^2}{k_0^2}\right )\, \frac{k_0}{2k}\, \ln
\frac {k_0+k}{k_0-k}
\right ]
\label{e4}
\eea
with the thermal SU(2) gauge boson mass $m_g^2=2g^2T^2/9$ and the SU(2)
indices $a,b$ of the adjoint representation. The remaining
two diagrams involving a hard scalar loop show the same momentum and energy
dependence as in (\ref{e4}) with a contribution to the thermal mass of
$g^2T^2/18$. In total the expression (\ref{e4}) is changed only by replacing
$m_g^2$ by $m_G^2=5g^2T^2/18$.

The effective gauge boson propagator in Coulomb gauge is given by
\bea
{D^{ab}_L}^\star (K) & \equiv & {D^{ab}_{00}}^\star =\frac{\delta ^{ab}}
{k^2-\Pi _L(K)}, \nonumber \\
{D^{ab}_T}^\star (K) & \equiv & \frac{1}{2} \left (\delta _{ij}-\frac{k_ik_j}
{k^2}\right )\> {D^{ab}_{ij}}^\star =\frac{\delta ^{ab}}{K^2-\Pi _T(K)}.
\label{e5}
\eea

The Higgs damping rate at zero momentum is defined by
\beq
\gamma _H(p=0)=-\frac{1}{4m_H}\> Im\,
\Xi _{\alpha \alpha} ^\star (\omega =m_H, p=0).
\label{e6}
\eeq
To leading order in the coupling constant $g$ the imaginary part of the
Higgs self energy $\Xi ^\star $ comes from the diagram in Fig.3. The damping
mechanism can be read off from this diagram by cutting the internal
lines \cite{WEL2}. The imaginary part of this diagram results from the
discontinuous part of the effective gauge boson propagator (Landau damping)
corresponding to the scattering of the soft Higgs off a thermal
gauge boson or Higgs via the exchange of a soft gauge boson. (There is
no pole-pole contribution due to kinematical reasons.) Owing to
the absence of effective vertices and an imaginary part in the effective
Higgs propagator there are no bremsstrahlung contributions, i.e.
$2\leftrightarrow 3$ processes, to the damping rate in leading order in
contrast to the much more complicated damping rate of a soft gluon \cite{BP2}.

Since we consider the damping
rate only at zero momentum, the transverse (magnetic) part of the
gauge boson propagator does not contribute, leading to an infrared
finite result due to Debye screening in the effective longitudinal
(electric) propagator.

Using the Matsubara formalism we find
\beq
\gamma _H(0)=\frac{3}{32\pi }\, g^2\, T\> \int _0^\infty dk\; \frac
{(\omega_k+m_H)^3}{\omega _k^2}\> \rho _L(\omega \!=\! \omega _k\! -\!
m_H,k),
\label{e7}
\eeq
where $\omega _k^2=k^2+m_H^2$ and
\beq
\rho _L(\omega ,k)=- \frac{1}{\pi}\> Im\, D^\star _L(\omega ,k)
\label{e8}
\eeq
is the discontinuous part of the longitudinal gauge boson spectral function
\cite{PI1}.

The remaining integral can be solved only numerically leading to Fig.4, where
the Higgs damping rate $\gamma _H/g^2T$ is shown as a function of
$\lambda /g^2$. The damping rate depends only weakly on the Higgs self
interaction coupling $\lambda $, increasing from 0.018 $g^2T$ at $\lambda
=0$ to 0.029 $g^2T$ if $\lambda $ tends to infinity.

The damping rate of the gauge boson at zero momentum is identical to the
one calculated by Braaten and Pisarski \cite{BP2}. There is no
contribution from the polarization diagram by a Higgs pair due to
kinematical reasons. Furthermore the final result is independent of
the thermal gauge boson mass $m_G$. (Hence the gluon damping rate
does not depend on the number of quark flavors in the quark-gluon
plasma.) Therefore the gauge boson damping rate is given by the
result in Ref.\cite{BP2} with $N_c=2$, i.e.
\beq
\gamma _{L,T}(0)=0.176\> g^2\, T.
\label{e9}
\eeq
(The damping rates of a longitudinal and a transverse gauge boson are
identical at zero momentum.) It should be noted that the gauge boson
damping rate is about an order of magnitude larger than the one of
a Higgs particle. This is caused partly by the fact that there is no
bremsstrahlung contribution in the Higgs damping and partly by different
SU(2) group factors.

Finally we would like to consider the damping rate of a hard Higgs particle
with a momentum of the order of the temperature or larger. Although it is not
related to the classical Lyapunov exponent, it might be of interest for
cosmological problems, since it determines the behaviour of thermal
and energetic Higgs particles, e.g. relaxation times and energy loss, in
an electroweak plasma in or near equilibrium.

The hard damping rate follows from the same diagram (Fig.3) as the soft one.
However, now it is sufficient to consider a bare Higgs propagator, i.e.
setting $m_H=0$. Analogously to the case of a hard quark damping rate
\cite{TG} we obtain an logarithmically infrared divergent result due to
the absence of screening in the transverse part of the effective gauge boson
propagator. Assuming an infrared cutoff of the order $g^2T$, as for example
a magnetic screening mass \cite{PI2} or by the Bloch-Nordsieck mechanism
\cite{BI} we find to logarithmic accuracy
\beq
\gamma _H(p{\buildrel >\over \sim}T)=\frac{3g^2T}{16\pi }\> \ln \frac{1}{g}.
\label{e10}
\eeq

\newpage
\section{Lattice simulation}

Now we turn to the computation of the Lyapunov exponent of the coupled
Yang-Mills Higgs system starting from a classical real-time lattice
description.
The Hamiltonian of the continuum Higgs model,
\beq
H = \frac{1}{2} E^a_i E^a_i + 
 \frac{1}{4} F_{ij}^a F^{ij}_a + 
 \dot{\Phi}^\dag \dot{\Phi} + (D_i \Phi )^\dagger \, (D_i \Phi )
 - \mu ^2  \Phi ^\dag  \Phi +  \lambda \,  (\Phi ^\dagger \, \Phi )^2,
\eeq
includes the complex Higgs doublet,
\beq
\Phi \, = \, \frac{1}{\sqrt{2}}
\left( 
\begin{array}{c} -\Phi_1+i\Phi_2 \\ \Phi_3+i\Phi_0 \end{array}
\right)
\eeq
which can be represented by four real components also as a
quaternion
\beq
\Phi  \, = \, \frac{1}{\sqrt{2}} 
\left( \Phi_0 + i\tau^c \Phi^c \right)
\eeq
with $c=1,2,3$ and the $\tau$-s being the Pauli-matrices.
These representations are related via the scalar product
of SU(2) group elements 
\hbox{$\langle A, B \rangle = \frac{1}{2}$tr$(AB^\dag)$}
\beq
\Phi^\dag\Phi = \frac{1}{2} \left(
\Phi^2_0 + \Phi^2_1 + \Phi^2_2 + \Phi^2_3 \right)
= \frac{1}{2} \langle \Phi, \Phi \rangle .
\eeq
The Hamiltonian can be approximated on a spatial lattice
with lattice spacing $a$ by using the variables
\beq
\phi_x = \frac{ag}{2} \Phi(x)
\eeq
as quaternion variables on each lattice site $(x=0,\ldots,N^3-1)$
and
\beq
U_{x,i} = e^{i \frac{ag}{2} \tau^c A^c_i(x) }
\eeq
as unit length quaternions \hbox{$(\langle U,U \rangle = 1)$}
on each lattice link from $x$ to $x+i$, $(i=0,1,2)$.
The canonically conjugate momenta of these lattice variables
are
\beq
P_{x,i} = \frac{4a}{g^2} \dot{U}_{x,i}
\qquad {\rm and} \qquad
\psi_x = \frac{4a}{g^2} \dot{\phi}_x,
\eeq
where the dot signals time derivation. The lattice Hamiltonian is
\cite{BOOK}
\bea
H &=& \frac{g^2}{4a} \left(
\sum_x \limits \frac{1}{2} \langle \psi_x,\psi_x \rangle +
\sum_{x,i} \limits \frac{1}{2} \langle P_{x,i},P_{x,i} \rangle \right) 
\quad + \NL \NL
& & \frac{4}{g^2a} \left(
\sum_x \limits (3-\frac{m_f}{2})\langle \phi_x,\phi_x \rangle 
+ \frac{g_f}{4} \langle \phi_x,\phi_x \rangle^2 
+  \sum_{x,i} \limits 1 - 
\langle U_{x,i}, \frac{1}{4} V_{x,i} + \phi_x \phi_{x+i}^\dag \rangle \right)
\label{LHAM}
\eea
Here $m_f=a^2\mu^2$ and $g_f=4\lambda/g^2$ with $a$ being the lattice
spacing. We used the following lattice approximation of the covariant
derivative of the Higgs field
\beq
a \cdot (D_i\phi_x) = U_{x,i}\phi_{x+i} - \phi_x,
\label{LDER}
\eeq
and the identity
\beq
\sum_{x,i} \limits \langle U_{x,i}\phi_{x+i}, U_{x,i}\phi_{x+i} \rangle
= 
\sum_{x,i} \limits \langle \phi_x, \phi_x \rangle.
\eeq
Now scaling the time to $t^\prime = t/a$ and the energy to
$H^\prime = ag^2 H / 4$ leads to
\bea
H^\prime &=& \sum_x \limits
\frac{1}{2} \langle \psi_x,\psi_x \rangle +
(3 - \frac{m_f}{2}) \langle \phi_x,\phi_x \rangle +
\frac{g_f}{4} \langle \phi_x,\phi_x \rangle^2  \NL 
&+& \sum_{x,i} \limits
\frac{1}{2} \langle P_{x,i},P_{x,i} \rangle + 1 -
\langle U_{x,i}, \frac{1}{4}V_{x,i} + \phi_x\phi^\dag_{x+i} \rangle.
\eea
Here and above the complement link variable, $V_{x,i}$,
is constructed by adding triple products of group elements, $U$,
on oriented links closing elementary plaquettes with the
chosen link $(x,i)$ such that
\beq
\sum_{x,ij} \limits \frac{1}{2} \, {\rm tr} \, U_{x,ij} =
\sum_{x,ij} \limits \langle U_{x,ij},1 \rangle =
\frac{1}{4} 
\sum_{x,i}  \limits \langle U_{x,i},V_{x,i} \rangle.
\eeq
Here $U_{x,ij}$, the plaquette variable is the product of
the four link variables $U$ circumventing the plaquette
with corner $x$ and lying in the $ij$ plane.

The scaled equations of motion are
\bea
\dot{U}_{x,i} = P_{x,i} &\qquad & \dot{\phi}_x = \psi_x \NL \NL
\dot{P}_{x,i} = V_{x,i} + \phi_x\phi^\dag_{x+i} &\qquad &
\dot{\psi}_x = W_x \, - \, F_x
\eea
with
\beq
W_x = \sum_i \limits U_{x,i}\phi_{x+i} +
U^\dag_{x-i,i}\phi_{x-i}
\eeq
and
\beq
F_x = \left( 6 - m_f + g_f\langle \phi_x,\phi_x \rangle \right)
\phi_x.
\eeq
A gauge transformation which leaves the lattice Hamiltonian
(\ref{LHAM}) invariant and changes the lattice derivative
(\ref{LDER}) covariantly is given by
\bea
U_{x,i}^\prime &=& g_x^\dag \, U_{x,i} \, g_{x+i} \NL 
\phi_x^\prime  &=& g_x^\dag \, \phi_x
\eea
with $g_x$ being an arbitrary SU(2) group element.
The static Noether charge corresponding to such time independent
gauge transformations becomes
\beq
\Gamma^\prime_x = \frac{g^2}{4} \Gamma_x
= \sum_i \limits
\left( U^\dag_{x-i,i}P_{x-i,i} \, - \, P_{x,i}U^\dag_{x,i} \right)
- \psi_x \phi_x^\dag.
\eeq

In order to relate the numerical simulation to physical parameters
we consider the average energy density
\beq
\overline{\varepsilon} = \frac{1}{a^3N^3} \frac{4}{ag^2} H^\prime
= \frac{12}{a^4g^2} \left( \frac{1}{3N^3} H^\prime \right).
\eeq
The expression in the last brackets is the scaled energy per link
on an \hbox{$N\times N\times N$} periodic lattice.
Comparing it with the energy density of an ideal gas of
SU(2) gluons and Higgs bosons at a fixed temperature $T,$ 
i.e. with \hbox{$\epsilon = \pi^2T^4/3$} we arrive at
\beq
\frac{1}{3N^3} H^\prime = \frac{\pi^3}{9} \alpha_w (aT)^4.
\eeq
When obtaining the equivalent temperature
the Higgs fields are  also considered,
since -- as the results of the numerical
simulations -- the energy is equally partitioned
between electric magnetic and rest degrees of freedom in the
final, chaotized state.
$\alpha_w = g^2/4\pi$ is the weak fine structure constant
which can be renormalized at high temperature perturbatively.
Its known dependence on the physical temperature (via the
perturbative beta function) sets the physical scale in the
originally scaleless classical simulation.

Finally we note that in order to ensure the unitarity of the group
elements \hbox{$\langle U,U \rangle = 1$} and its conservation
\hbox{$\langle U,P \rangle = 0$} during the solution of the
equations of motion Lagrange multipliers must be used.
Including a small scaled timestep factor $dt/a=h$ into the
conjugate momenta and its square into the respective forces
we arrive at the following implicit recursion scheme,
which conserves the static Noether charge \cite{ALGO} 
\bea
U^\prime &=& U + (P^\prime - \varepsilon U) \NL
P^\prime &=& P + (V - \mu U + \varepsilon P^\prime) \NL
\phi^\prime &=& \phi + \psi^\prime \NL
\psi^\prime &=& \psi + (W - F)
\eea
with $\mu = \langle P^\prime,P^\prime \rangle$
and $\varepsilon = \langle U,P^\prime \rangle$.


\section{Lyapunov Exponents}

\vspace{0.5cm}
We use the following gauge invariant definition of the distance
between two different gluon and Higgs field configurations, respectively
\bea
D[U,U^*] &=& \sum_{x,i} \limits \left|
\langle U, V \rangle \, - \, \langle U^*, V^* \rangle \right|,
\NL 
D[\Phi,\Phi^*] &=& \sum_{x,i} \limits \left|
\langle \Phi, \Phi \rangle \, - \, \langle \Phi^*, \Phi^* \rangle \right|.
\eea
We sampled parallel runs on an $N=10$ cubic lattice of randomly
initialized $U,\Phi$ configurations with unit length.
In the dynamical simulation the length of the Higgs field
could change freely according to the dynamics while the length
of the gauge field remained equal to one. We repeated each calculation with
slightly rotated initial configuration $U^*,\Phi^*$.
The time evolution of these initially small distances has been
observed over a time long enough for the saturation.
(Distances in a compact space tend to saturate.)

As can be seen in Fig.5 the configurations diverge exponentially
in the $U$ as well as in the $\Phi$ configuration space.
In the linear range of this evolution the leading Lyapunov
exponent can be extracted simply
\beq
h = \frac{d}{dt} \ln D(t).
\eeq
Choosing maximally randomized purely magnetic gauge field configurations
and similar Higgs field configurations initially we studied the
chaotic behavior due to the extraction of leading Lyapunov exponents
for different values of $g_f=4\lambda/g^2$. 
As a result of this study the Higgs and gauge boson Lyapunov
exponents are plotted in Fig.6. As expected, the Lyapunov exponent
of the gauge boson does not depend on $\lambda $ and agrees with twice
of the damping rate (\ref{e9}) within the numerical accuracy.

Surprisingly the Higgs Lyapunov Exponent shows a behavior qualitatively and
quantitatively different from that of the damping rate obtained in the
resummed one loop approximation (Fig.4).
We remind, that this is contrary to 
the case of the pure gauge field system, where the Lyapunov exponent has 
been found to be twice of the gluon damping rate at zero momentum\cite{BGM95}.

The result of the classical lattice dynamics can be understood easily:
for very high values of $\lambda$ the $\Phi$-fields are constrained
to keep a rather definite value of $\langle\Phi,\Phi\rangle$.
The number of light degrees of freedom
is reduced from four to three, so the Higgs sector becomes 
less chaotic.
In the simulations presented here the gauge fields were initialized 
with random phases  covering the whole group SU(2) uniformly 
in the magnetic sector, so these degrees of freedom could not take any
energy from the Higgs fields over. 
The electric sector was able to receive
energy also from the Higgs, but because of the final equipartition with
the magnetic degrees of freedom 
the total energy converted into chaotic motion remained limited.

$\,$ From the results of the numerical simulation presented here we must
conclude that the Higgs damping rate, contrary to the gluon damping,
is not closely related to the chaotic behavior of the classical fields.
While the chaotic behavior of Higgs fields
is rather set by the damping rate of gluons which couple
to the Higgs covariantly and decreases somewhat at high
$\lambda$ values, their resummed damping rate at high temperature
increases with increasing Higgs-self
coupling due to the thermal Higgs-mass which sets the scale 
in that calculation (\ref{e7}).

\acknowledgements

M.H.T. thanks E. Braaten and A.K. Rebhan for helpful discussions.
This work was supported by the Deutsche Forschungsgemeinschaft
and the Hungarian Academy of Sciences (DFG-MTA 79/1995), by the
Hungarian National Scientific Research Fund (OTKA T014213) and
by BMBF and GSI. T.S.B. thanks for the fellowship of the
University of Bergen  which enabled him to visit
the Physics Department there and to carry out a part of the 
numerical calculations.

\begin{figure}

\centerline{\psfig{figure=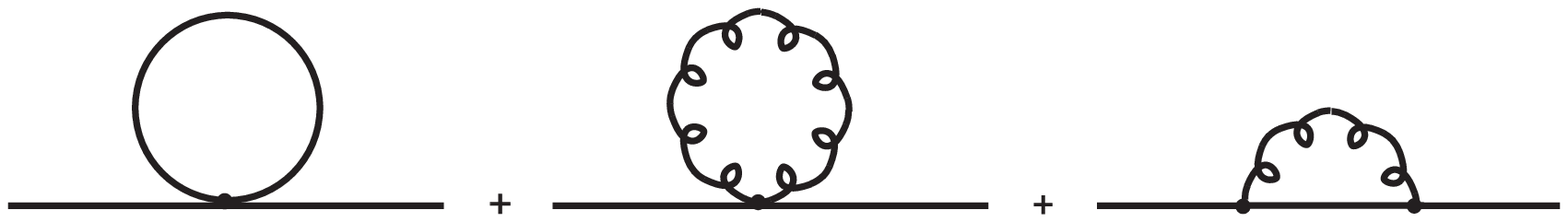,width=10.0cm}}

\caption{HTL contributions to the Higgs self energy.}
\end{figure}

\begin{figure}

\centerline{\psfig{figure=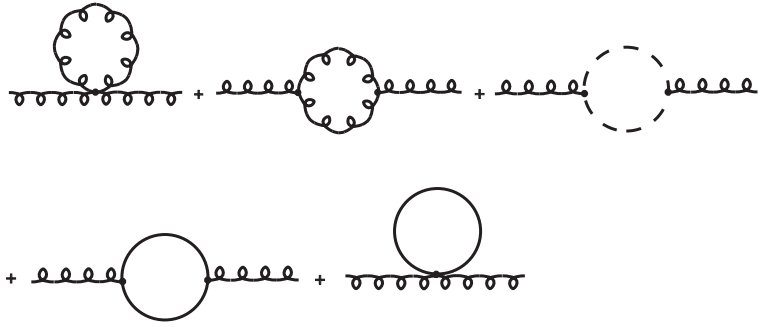,width=10.0cm}}

\caption{HTL contributions to the gauge boson self energy.}
\end{figure}

\begin{figure}

\centerline{\psfig{figure=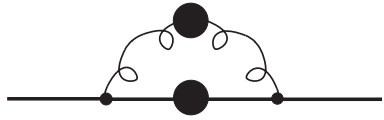,width=5.0cm}}

\caption{Higgs self energy contribution to the damping rate.}
\end{figure}

\begin{figure}

\centerline{\psfig{figure=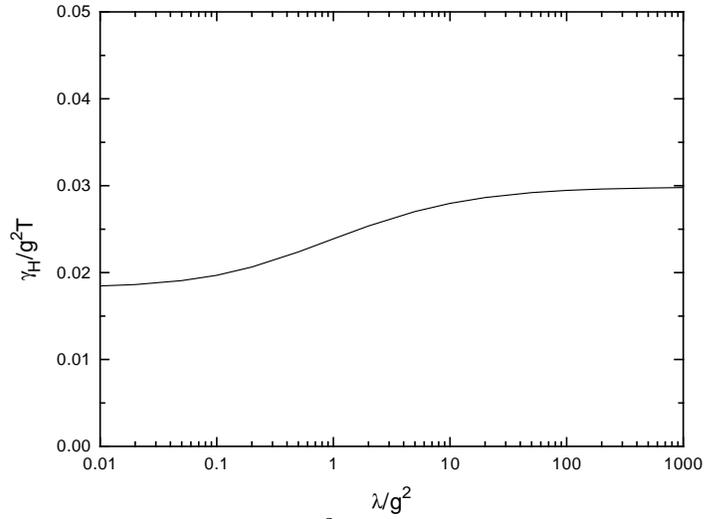,width=10.0cm}}

\caption{Higgs damping rate $\gamma _H/g^2T$ at zero momentum as a
function of $\lambda /g^2$.}
\end{figure}

\begin{figure}

\centerline{\psfig{figure=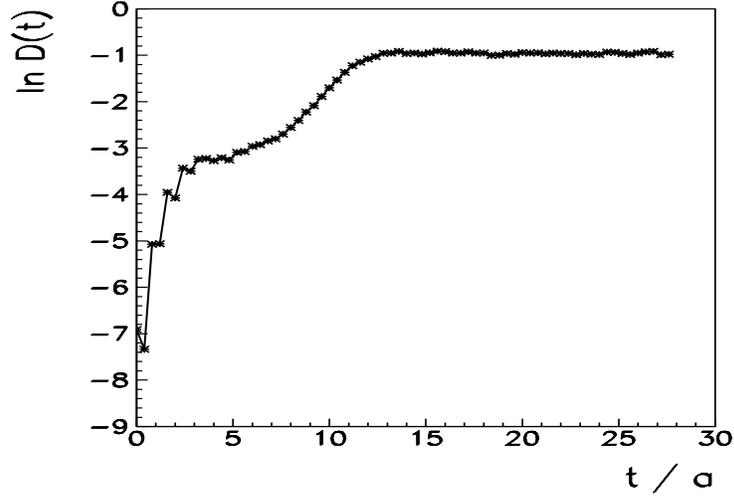,width=9.00cm,height=9.00cm}}

\caption{Time evolution of the distance of two initially adjacent
	Higgs field configurations at $4\lambda/g^2=1024$.}
\end{figure}

\begin{figure}

\centerline{\psfig{figure=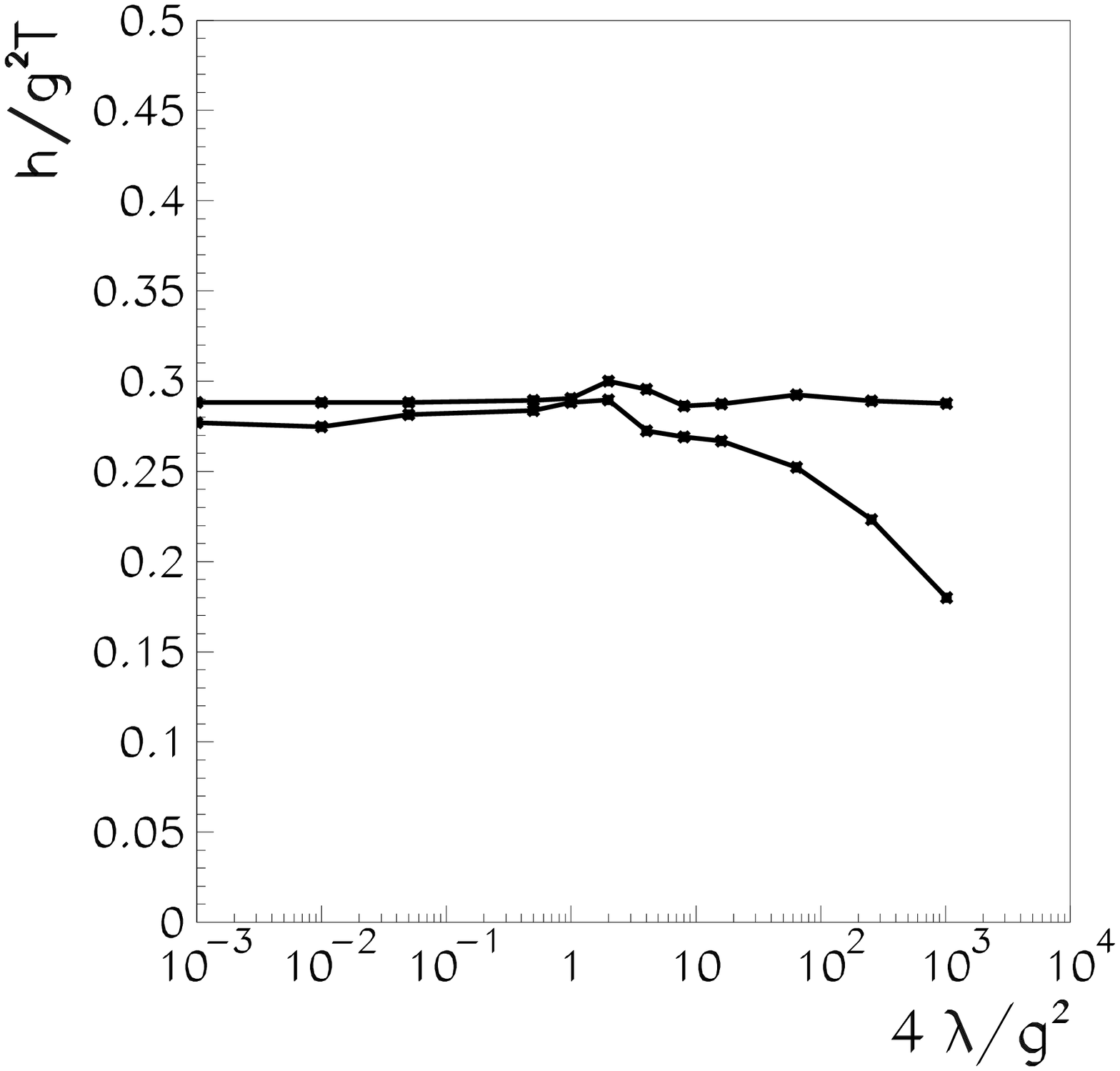,width=9.00cm,height=9.00cm}}

\caption{The gauge field (upper curve)  and Higgs field (lower curve) 
	leading Lyapunov exponents
        as a function of $4\lambda/g^2$ at high temperature. The numerical
        results have an error bar of about $20 \%$.}
\end{figure}

\begin{references}

\bibitem{BP1}
E. Braaten and R.D. Pisarski, Nucl. Phys. {\bf B337}, 569 (1990).

\bibitem{BP2}
E. Braaten and R.D. Pisarski,  Phys. Rev. D {\bf 42}, 2156 (1990).

\bibitem{BGMT}
T.S. Bir\'o, C. Gong, B. M\"uller, and A. Trayanov, Int. Jour. Mod. Phys.
C {\bf 5}, 113 (1994).

\bibitem{BGM95}
T.S.Bir\'o, C.Gong, and B.M\"uller, Phys. Rev. D {\bf 52}, 1260 (1995).

\bibitem{KRS}
U. Kraemmer, A.K. Rebhan, and H. Schulz, Ann. Phys. (N.Y.) {\bf 238}, 286
(1995).

\bibitem{THO}
M.H. Thoma, Z. Phys. C {\bf 66}, 491 (1995).

\bibitem{IZ}
C. Itzykson and J.-B. Zuber, {\it Quantum Field Theory} (McGraw-Hill,
New York, 1980).

\bibitem{WEL1}
H.A. Weldon, Phys. Rev. D {\bf 26}, 1394 (1982).

\bibitem{KLI}
V.V. Klimov, Zh. Eksp. Teor. Fiz. {\bf 82}, 336 (1982) [Sov. Phys. JETP
{\bf 55}, 199 (1982)].

\bibitem{WEL2}
H.A. Weldon, Phys. Rev. D {\bf 28}, 2007 (1983).

\bibitem{PI1}
R.D. Pisarski, Physica {\bf A158}, 146 (1989).

\bibitem{TG}
M.H. Thoma and M. Gyulassy, Nucl. Phys. {\bf B351}, 491 (1991).

\bibitem{PI2}
R.D. Pisarski, Phys. Rev. Lett. {\bf 63}, 1129 (1989).

\bibitem{BI}
J. Blaizot and E. Iancu, Saclay Report No. T95/146, 1995 (unpublished).

\bibitem{BOOK}
T.S. Bir\'o, S.G.Matinyan, and  B. M\"uller,
{\it Chaos and Gauge Field Theory} (World Scientific, Singapore, 1995).

\bibitem{ALGO}
T.S. Bir\'o, Int. J. Mod. Phys. C {\bf 6}, 327 (1995).


\end{references}
\end{document}